\documentclass[12pt]{article}

\usepackage[T1]{fontenc}
\usepackage{mathbbol,bbm}
\usepackage{amsfonts}
\usepackage[latin2]{inputenc}
\usepackage{t1enc}
\usepackage{amssymb}
\usepackage{amsmath}
\usepackage[mathscr]{eucal}
\usepackage{amsthm}
\usepackage{array}
\usepackage{color}

\theoremstyle{definition} 
\newtheorem{defin}{Definition}[section]
\newtheorem{thm}[defin]{Theorem}

\newtheorem{ex}[defin]{Example}
\newtheorem{cor}[defin]{Corollary}

\newtheorem{prop}[defin]{Proposition}
\newtheorem{Lemma}[defin]{Lemma}
\newtheorem{lemma}[defin]{Lemma}
\newtheorem{remark}[defin]{Remark}

\def\bz{\left(}
\def\jz{\right)}
\def\Bz{\big(}
\def\Jz{\big)}
\def\ki{\textit}
\def\vfi{\varphi}
\def\ffi{\varphi}
\def\half{\frac{1}{2}}
\def\ep{\varepsilon}
\def\F{\mathcal{F}}
\def\X{\mathcal{X}}
\def\Y{\mathcal{Y}}

\def\B{\mathcal{B}}
\def\iH{\mathcal{H}}
\def\iK{\mathcal{K}}
\def\egy{\mathbf{\mathrm{1}}}

\def\A{\mathcal{A}}
\def\B{\mathcal{B}}
\def\C{\mathcal{C}}

\def\Ee{\mathcal{E}}
\def\Eu{\mathcal{E}_{\unit}}
\def\Et{\mathcal{E}_{e^{ta}}}

\def\kii{\textit}
\def\kiii{}
\def\sa{^{\mathrm{sa}}}
 
\def\bC{\mathbb{C}}
\def\bN{\mathbb{N}}
\def\inta{\Phi}

\def\loc{_{\Lambda}}

\def\sa{^{\mathrm{sa}}}
\def\aut{\alpha}

\def\Pn{\mathcal{P}_0}
\def\alg{\C}
\def\algloc{\C_{\mathrm{loc}}}
\def\Q{Q_{m}}
\def\obs{a}
\def\cp{_{\mathrm{CP}}}
\def\bbeta{\eta}
\def\d{\mathrm{d}}
\def\state{\vartheta}

\newcommand{\norm}[1]{\left\| #1\right\|}
\newcommand{\inner}[2]{\langle #1 , #2\rangle}
\newcommand{\abs}[1]{\left| #1 \right|}

\newcommand{\vect}[1]{\mathbf{#1}}

\newcommand{\diad}[2]{|#1\rangle\langle #2|}
\newcommand{\pr}[1]{\diad{#1}{#1}}
\newcommand{\D}{\hat}
\newcommand{\sr}[2]{S\bz #1 , #2\jz}
\newcommand{\msr}[2]{s\bz #1 , #2\jz}
\newcommand{\pressure}[2]{p_{#1}(#2)}
\newcommand{\momgen}[2]{f_{#1,#2}}
\newcommand{\chb}[3]{C_{#1}\bz #2 , #3\jz}
\newcommand{\pmin}[2]{P_{\mathrm{min}}(#1 : #2)}
\newcommand{\pn}[2]{P_{A_n}(#1 : #2)}
\newcommand{\sequence}[1]{\vec{#1}}

\newcommand{\N}{\mathbb{N}}
\newcommand{\Z}{\mathbb{Z}}
\newcommand{\R}{\mathbb{R}}

\newcommand{\iC}{\mathbb{C}}
\newcommand{\s}{\mbox{ }}
\newcommand{\ds}{\mbox{ }\mbox{ }}

\newcommand{\unit}{\Eins}

\DeclareMathOperator{\id}{id}

\DeclareMathOperator{\tmap}{\Phi}
\DeclareMathOperator{\tmapp}{\Psi}

\DeclareMathOperator{\Tr}{Tr}

\DeclareMathOperator{\exposed}{ex}
\DeclareMathOperator{\diam}{diam}
\DeclareMathOperator{\di}{d}
\DeclareMathOperator{\supp}{supp}

\DeclareMathOperator{\spect}{spec}
\DeclareMathOperator{\mg}{\Gamma}

\begin{document}

\centerline{\huge Large deviations and Chernoff bound for certain}
\medskip

 \centerline{\huge correlated states on a spin chain}
\bigskip
\s

\bigskip

 \centerline{\large Fumio Hiai\footnote{Electronic mail: hiai@math.is.tohoku.ac.jp}, 
Mil\'an Mosonyi\footnote{Electronic mail: milan.mosonyi@gmail.com}}
  \medskip

\centerline{\textit{Graduate School of Information Sciences, Tohoku University}}

\centerline{\textit{Aoba-ku, Sendai 980-8579, Japan}}
\bigskip

\centerline{\large Tomohiro Ogawa\footnote{Electronic mail: ogawa@quantum.jst.go.jp}}
\medskip

\centerline{\textit{PRESTO, Japan Science and Technology Agency}}

\centerline{\textit{4-1-8 Honcho Kawaguchi, Saitama, 332-0012, Japan}}
\bigskip
\s

\medskip

\begin{abstract}
In this paper we extend the results of Lenci and Rey-Bellet on the large deviation upper bound of the distribution measures of local Hamiltonians with respect to a Gibbs state, in the setting of translation-invariant finite-range interactions. We show that a certain factorization property of the reference state is sufficient for a large deviation upper bound to hold and that this factorization property is satisfied by Gibbs states of the above kind as well as finitely correlated states. As an application of the methods the Chernoff bound for correlated states with factorization property is studied.
In the specific case of the distributions of the ergodic averages of a one-site observable with respect to an ergodic finitely correlated state the spectral theory of positive maps is applied to prove the full large deviation principle. 
\medskip

 
\end{abstract}

\section{Introduction}

 Several of the main results of classical probability theory have been generalized to the quantum setting up till now. The main difference from the classical case is the use of non-commutative algebras, and though the proofs usually rely on their classical equivalents, new methods are required to circumvent the difficulty arising from the non-commutativity of  observable quantities. In this paper we investigate the large deviation principle (LDP) and a strongly related subject, the Chernoff bound for symmetric hypothesis testing for a class of states on a spin chain, extending the results of \cite{Aud,LRB,NR,SzN}.

Following the approach of \cite{LRB} and \cite{NR} we study the existence of the pressure function of a finite-range translation-invariant interaction $\inta$ with respect to a translation-invariant state $\omega$ on a spin chain. We show that a sufficient condition for the pressure to exist is a certain (upper) factorization property of the state $\omega$, that for Gibbs states follows from the results of \cite{LRB} on perturbations of Gibbs states. The existence of the pressure yields the existence of the asymptotic logarithmic moment generating function in the theorem of G\"artner and Ellis, which in turn implies the large deviation upper bound for the distributions of the local Hamiltonians of $\inta$ with respect to the state $\omega$. We show that finitely correlated states \cite{FNW} satisfy this upper factorization property, therefore our results extend the large deviation bounds obtained in \cite{LRB} to the class of finitely correlated states (Section \ref{existence}).

In \cite{NR} the full LDP was proven for the distributions of the ergodic averages of a one-site observable with respect to a high-temperature KMS (Gibbs) state, using a cluster expansion technique. In this paper we prove the same result for ergodic finitely correlated states, using the spectral theory of positive operators on finite-dimensional $C^*$-algebras (Section \ref{fcLDP}).

The Chernoff bound gives the exponential decay rate of the average of the error probabilities of the first and the second kinds when a binary test is performed on an increasing number of copies of the same system, either all prepared in some given state $\omega$ or another state $\sigma$. The quantum version of the corresponding classical theorem has recently been proven in \cite{SzN} (lower bound) and \cite{Aud} (upper bound). In Section \ref{chernoff section} we investigate the generalized situation when state discrimination is performed between two states of an infinite chain, by making binary measurements on increasing finite-size restrictions. Relying on the results of \cite{Aud,SzN} we prove the Chernoff upper bound for states satisfying an upper factorization property (hence for all finitely correlated states) and both upper and lower bounds for states satisfying both upper and lower factorization properties (like Gibbs states). 

To make the text more self-contained we include a section on preliminaries on quantum spin chains and large deviation theory and three appendices; on the perturbation results of Gibbs states developed in \cite{LRB} (Appendix \ref{gibbs}); on the construction of finitely correlated states (Appendix \ref{fcs}) and on the spectral theory of positive maps (Appendix \ref{pos}). 
A short introduction to hypothesis testing is included in Section \ref{chernoff section}.
 
For further results on large deviations in quantum systems we refer to \cite{Bjelakovic2,BLP,GLM,HMOP,LLS,Petz2,PRV}. In Section \ref{concluding remarks} we point out some connections to the works  \cite{Bjelakovic} and \cite{HMOP}. 

\section{Preliminaries}
\subsection{Quantum spin chains}

Let $\iH$ be a finite-dimensional Hilbert space and $\A\subset\B(\iH)$ be a $C^*$-subalgebra of $\B(\iH)$. The infinite spin chain with one site-algebra $\A$ is $\alg:=\otimes_{k\in\Z}\A$, which is the $C^*$-inductive limit of the algebras $\{\otimes_{k\in\Lambda}\,\A\,:\,\Lambda\in\Pn\}$ under the natural inclusions
\begin{equation*}
 \iota_{\Lambda',\Lambda}:\,\otimes_{k\in\Lambda}\A\hookrightarrow\otimes_{k\in\Lambda'}\A\,,\ds\ds\ds 
 a\mapsto a\otimes\bz \otimes_{k\in\Lambda'\setminus\Lambda}\unit_{\A}\jz\,,
 \end{equation*}
where $\Pn:=\{\Lambda\subset \Z\,:\,0<|\Lambda|<\infty\}$. We use the notation $\alg_{\Lambda}$ for the embedded image of $\otimes_{k\in\Lambda}\A$ into $\alg$, and usually identify the two. The \ki{algebra of local observables} is defined as $\algloc:=\bigcup_{\Lambda\in\Pn}\alg_{\Lambda}$. 
 
 Positive linear functionals $\omega:\,\alg\to\iC$ satisfying $\omega(\unit)=1$ (where $\unit$ is the unit of $\alg$) are called states. 
The distribution of a local observable $a\in\alg_{\Lambda}\sa:=\{x\in\alg_{\Lambda}:x^*=x\}$ ($\Lambda$ finite) with respect to a state $\omega$ is the probability measure
\begin{equation*}
\mu_{a}(B):=\omega\bz\egy_B(a)\jz\,;\ds\ds\ds B\subset\R\,,
\end{equation*}
where $\egy_B(a)\in\alg_{\Lambda}$ is the spectral projection of $a$ corresponding to the set $B$. Using functional calculus and the Riesz-Markov theorem \cite{RS}, one can also define the distribution of an arbitrary (not necessarily local) observable through
\begin{equation*}
\int\,g\,d\mu_{a}:=\omega\bz g(a)\jz\,;\ds\ds\ds g:\,\spect(a)\to\iC\ds\mathrm{continuous}\,.
\end{equation*}
The interpretation of the distribution of $a$ in quantum mechanics is that the measurement result falls into the set $B$ with probability $\mu_a(B)$ when the observable $a$ is measured in the state $\omega$ of the system.

The maps $\gamma_{n,m}:\,\otimes_{k\in [n,m]}\A\to\otimes_{k\in[n,m+1]}\A\,,\ds a\mapsto \unit_{\A}\otimes a$ ($n<m$) extend uniquely to the \ki{shift automorphism} of $\alg$. A state $\omega$ is called \ki{translation-invariant} if $\omega\circ\gamma=\omega$. The set of all translation-invariant states is convex; its extremal points are called \ki{ergodic states}. The restrictions
$\omega_{\Lambda}(a):=\omega(a),\s a\in\alg_{\Lambda}$ of a translation-invariant state satisfy the compatibility relations
\begin{equation*}
\omega_{[1,n+1]}(a\otimes\unit_{\A})=\omega_{[1,n+1]}(\unit_{\A}\otimes a)=\omega_{[1,n]}(a)\,;\ds\ds\ds a\in\C_{[1,n]}.
\end{equation*}
On the other hand, any sequence of states $\{\omega_n\,:\,n\in\N\}$ on the algebras $\alg_{[1,n]}$ that satisfies the above compatibility relations extends uniquely to a translation-invariant state $\omega$ with $\omega_{[1,n]}=\omega_n$. 
The \ki{density operator} $\D{\omega}_{\Lambda}$ of a local restriction $\omega_{\Lambda}$ is the unique element in $\C_{\Lambda}$ that satisfies $\rho(a)=\Tr_{\Lambda}\D{\rho}_{\Lambda}a,\s a\in\C_{\Lambda}$.

The above construction provides a mathematical model in statistical physics to describe identical finite-level systems located at the sites of the one-dimensional lattice $\Z$. The interaction between the systems localized at each site is described by a function $\inta:\,\Pn\to\alg\sa$ satisfying 
$\inta(X)\in\alg_X,\s X\in\Pn$\,. We will always assume that $\inta$ has \ki{finite range}, i.e.
\begin{equation*}
\di(\inta):=\inf\{d\,:\,\inta(X)=0\ds\text{when}\s\diam(X)>d\}<\infty
\end{equation*}
 and \ki{bounded surface energy}:
\begin{equation*}
\sup_{n\in\N}\,\norm{\sum\big\{\inta(X)\,:\,X\cap [-n,n]\ne\emptyset,\s X\cap (\Z\setminus[-n,n])\ne\emptyset \big\}}<\infty\,.
\end{equation*}
 The interaction $\inta$ is called translation-invariant if $\inta(X+1)=\gamma\bz\inta(X)\jz$. Obviously, a translation-invariant finite-range interaction automatically has bounded surface energy.
 
For each $\Lambda\in\Pn$ the interaction defines a \ki{local Hamiltonian} by $H_{\Lambda}:=\sum_{X\subset \Lambda}\,\inta(X)$ with corresponding \ki{local dynamics} $\aut^{\Lambda}$ and \ki{local Gibbs state} $\vfi^{\Lambda}_{G}$ by
\begin{equation*}
\aut^{\Lambda}_t(a):=e^{itH_{\Lambda}}\,a\,e^{-itH_{\Lambda}}\,,\ds\ds\ds \vfi^{\Lambda}_G(a):=\tau(e^{-H_{\Lambda}}a)/\tau(e^{-H_\Lambda})\,;\ds\ds
a\in\alg\, ,
\end{equation*}
with 
\begin{equation}\label{trace}
\tau:=\otimes_{k\in \Z}\bz\frac{1}{\dim\iH}\Tr\jz\,,
\end{equation}
 where $\Tr$ is the usual trace functional on $\B(\iH)$.
The restrictions to $\alg_{\Lambda}$ are denoted by $\aut_{\Lambda,t}$ and $\vfi_{\Lambda,G}$, respectively; i.e.
\begin{equation*}
\aut_{\Lambda,t}(a)=e^{itH_{\Lambda}}\,a\,e^{-itH_{\Lambda}}\,,\ds\ds\ds \vfi_{\Lambda,G}(a):=\Tr_{\Lambda}(e^{-H_{\Lambda}}a)/\Tr_{\Lambda}(e^{-H_{\Lambda}})\,;\ds\ds
a\in\alg_{\Lambda}\, .
\end{equation*}
For simplicity, we use the notation $\vfi_{n,G}$ for $\vfi_{[1,n],G}$.
Finite range condition together with bounded surface energy guarantees that the (thermodynamical) limits
\begin{equation*}
\vfi:=(\mathrm{weak}^*)\lim_{\Lambda\nearrow \Z}\vfi^{\Lambda}_G\,,\ds\ds\ds
\aut_t:=(\mathrm{strong})\lim_{\Lambda\nearrow \Z}\aut^{\Lambda}_t\,;\ds\ds t\in\R 
\end{equation*}
exist; $\aut$ is called the \ki{global dynamics} and $\vfi$ the \ki{global Gibbs state}, which is the unique KMS state for $\aut$ (see e.g.~\cite{Ar}, \cite[Proposition 6.2.47]{BR2} and \cite[Proposition 4.1.9]{Sakai}). Moreover, for a translation-invariant interaction any local element $a\in\algloc$ is an analytic element of the global automorphism group, i.e. $f_a(t):=\aut_t(a)$ has an analytic extension
$\aut_z(a)$ to the whole complex plain (see \cite{Araki,Araki2}).
Note that $\vfi_{[1,n]}\ne \vfi_{n,G}$ except for trivial cases. However, the following holds 
\cite{Araki,Araki2,ArIon}:
\begin{Lemma}\label{Araki}
There is a constant $\lambda>0$ such that
\begin{equation*}
\lambda^{-1}\vfi_{[1,n]}\le\vfi_{n,G}\le\lambda\,\vfi_{[1,n]}
\quad\mbox{for all $n\in\bN$}.
\end{equation*}
\end{Lemma}
 
\subsection{Large Deviation Principle}

Large deviation results describe the asymptotic behavior of probabilities of ``rare'' events.  
As a simple example, one can consider the behaviour of the average of a sequence of i.i.d. random variables taking values in a finite subset of $\R$. Without loss of generality one can assume the probability space to be $(\X^{\infty},\F,\omega)$, where $\X$ is a finite set, $\F$ is the $\sigma$-field generated by the cylinder sets and $\omega=\omega_1^{\otimes \infty}$ is a translation-invariant product probability measure. The shift on infinite sequences $\vect{x}\in\X^{\infty}$ is defined by $(S\,\vect{x})_k:=\vect{x}_{k+1}$ and it generates a shift $\gamma$ of functions $a:\,\X^{\infty}\to\iC$ through $\gamma(a):=a\circ S$. Now if $a$ depends on only one site, i.e. $a(\vect{x})=f(x_1)$ for some $f:\,\X\to\R$ then $a,\gamma(a),\gamma^2(a),\ldots$ is a sequence of i.i.d.~random variables with common expectation value $m$, and by the law of large numbers $\mu_n\bz [c,d]\jz\to 0$ as $n\to\infty$ for any non-degenerate interval 
$[c,d]$ that does not contain $m$, where $\mu_n$ is the distribution of $\frac{1}{n}\,\sum_{k=0}^{n-1}\,\gamma^k(a)$. Cram\'er's large deviation result \cite{DZ}
gives more detailed information on the speed with which these probabilities tend to $0$; it states that
\begin{equation*}
\frac{1}{n}\,\log\mu_n\bz [c,d]\jz\xrightarrow[n\to\infty]{}-\inf_{x\in [c,d]} I(x)\, ,
\end{equation*}
where $I(x):=\sup\{tx-F(t)\,:\,t\in\R\}$ is the Legendre-Fenchel transform of the logarithmic moment generating function 
\begin{equation*}
F(t):=\log\int\,e^{tx}\,d\mu_1(x)=\log\int\,e^{ta}\,d\omega\,.
\end{equation*}

In a more general context,  
 a sequence $(\mu_n)_{n\in\N}$ of probability measures on the Borel sets of $\R$ is said to satisfy the \ki{large deviation principle (LDP)} with rate function $I$ if 
\begin{equation*}
-\inf_{x\in\mathrm{int}\, H} I(x)\le \liminf_{n\to\infty}\,\frac{1}{n}\,\log\mu_n(H)\le\limsup_{n\to\infty}\,\frac{1}{n}\,\log\mu_n(H)\le -\inf_{x\in\overline H} I(x)
\end{equation*}
holds for any measurable $H\subset \R$, where $\overline{H}$ is the closure of $H$ and $\mathrm{int}\,H$ is its interior.
We say that $(\mu_n)_{n\in\N}$ satisfies the large deviation lower (resp.~upper) bound if the first (resp.~the last) inequality holds in the above chain.
 The rate function $I$ is called a \ki{good rate function} if the level sets $\{x\in\R\,:\,I(x)\le c\}$ are compact for any $c\in\R$.
A fundamental result in the theory of large deviations is a generalization of Cram\'er's theorem to the non-i.i.d.~setting by G\"artner and Ellis (see e.g.~\cite{DZ}). 
Before stating this result we recall that $y\in\R$ is called an \ki{exposed point} of the convex function $f:\R\to\R$ if for some $t\in\R$ the function $x\mapsto tx-f(x)$ has a strict maximum at $y$. We denote the set of exposed points by $\exposed(f)$.
\begin{thm} (G\"artner \& Ellis)\ds
Let $(\mu_n)_{n\in\N}$ be a sequence of probability measures on the Borel sets of $\R$ and let 
\begin{equation}\label{momgen0}
\mg_n(t):=\int\,e^{ntx}\,d\mu_n(x)\,;\ds\ds\ds t\in\R.
\end{equation}
If the limit 
\begin{equation*} 
 F(t):=\lim_{n\to\infty}\,\frac{1}{n}\,\log\,\mg_n(t)
\end{equation*}
 exists for all $t$ as an extended real number and is finite in a neighbourhood of $0$ then
\begin{align}
-\inf_{x\in\mathrm{int}\, H\,\cap\, \exposed(I)} I(x)&\le \liminf_{n\to\infty}\,\frac{1}{n}\,\log\mu_n(H)\label{lb}\\
&\le\limsup_{n\to\infty}\,\frac{1}{n}\,\log\mu_n(H)\le -\inf_{x\in\overline H} I(x)\nonumber
\end{align}
holds with the good rate function
 \begin{equation*} 
I(x)=\sup\{tx-F(t)\,:\, t\in\R\}\,.
\end{equation*}
If, moreover, $\mg$ is differentiable 
then the lower bound in (\ref{lb}) can be replaced with
\begin{equation*}
-\inf_{x\in\mathrm{int}\, H} I(x)\le \liminf_{n\to\infty}\,\frac{1}{n}\,\log\mu_n(H) \,.\label{lb2}
\end{equation*}
\end{thm}

Note that the rate function $I$ above is convex, as is always the case when the G\"artner-Ellis theorem is applied to derive the large deviation principle, hence by a rate function we will always mean a convex one.
\smallskip
 
Now if $\omega$ is a fixed state of a spin chain $\C$ and $\obs\in\alg_{[1,r]}$ is a local observable then one can use the G\"artner-Ellis theorem to study the large deviation properties of the sequence $(\mu_{n,a})_{n\in\N}$, where $\mu_{n,a}$ is the distribution of $\bar\obs_n:=\frac{1}{n}\,\sum_{k=0}^{n-1}\,\gamma^k(\obs)$ with respect to $\omega$. Apart from being formulated on a possibly non-commutative spin chain, this question generalizes the setting of Cram\'er's theorem in two directions: first, $\omega$ is not necessarily a product state, which is equivalent to the random variables being correlated; second, $\obs$ may be a multi-site observable, which corresponds to considering averages of the form
$\frac{1}{n}\sum_{k=0}^{n-1}\,f(X_{k+1},\ldots,X_{k+r})$ in the classical case. As a further generalization, one can replace $\bar\obs_n$ with $\frac{1}{n}H_{[1,n]}$, where $H_{[1,n]}$ is the local Hamiltonian of a translation-invariant finite-range interaction $\inta$. The large deviation properties of the corresponding distribution measures $\mu_{n,\inta}$ were studied in \cite{LRB}. Note that the previous example corresponds to
\begin{equation*}
\inta(X):=\begin{cases} \gamma^k(\obs) & \s \text{ if }\s X=[k+1,k+r] \text{ for some } k\,;\\
                        0 & \s \text{ otherwise}
\end{cases}
\end{equation*}
and $\mu_{n,a}=\mu_{n+r-1,\inta}$ for the interaction above.
 
The moment generating function of $\mu_{n,\inta}$ is
\begin{equation*}
\mg_n(t)=\int\,e^{ntx}\,d\mu_{n,\inta}(x)=\omega\Bz e^{tH_{[1,n]}}\Jz\,;\ds\ds\ds t\in\R\,.
\end{equation*}
Note that if $\omega=\tau$ is the trace state of (\ref{trace}) then
\begin{equation*}
\frac{1}{n}\log\mg_n(t)=\frac{1}{n}\log\Tr e^{tH_{[1,n]}}-\log\dim\iH\,,
\end{equation*}
and the first term is known to converge to $P(-t\inta)$, where $P$ is the \ki{pressure} (or mean free energy) functional \cite{Israel}. This motivates to use the term pressure for the limit 
\begin{equation}\label{pressure}
\pressure{\omega}{\inta}:=\lim_{n\to\infty}\,\frac{1}{n}\log\omega\Bz e^{-H_{[1,n]}}\Jz
\end{equation}
whenever it exists.

\section{LDP for finitely correlated states}\label{fcLDP}

We use the notations and results of Appendices \ref{fcs} and \ref{pos}. In particular, we denote a generating triple of a finitely correlated state $\omega$ by $(\B,\Ee,\rho)$, where $\rho$ is assumed to be faithful, and for a one-site observable $a$ we define the map
\begin{equation*}
\Et:\,\B\to\B\,, \ds\ds\ds \Et:\,b\mapsto\Ee\bz e^{ta}\otimes b\jz\,.
\end{equation*}

\begin{thm}\label{fcldp}
Let $\omega$ be an ergodic finitely correlated state, $\obs\in\alg_{\{1\}}\sa$ be a one-site observable and 
\begin{equation*}
\mu_{n,\obs}(B):=\omega\bz\egy_B\bz  \bar\obs_n\jz\jz\,;\ds\ds\ds B\subset \R
\end{equation*}
be the distribution of $\bar\obs_n:=\frac{1}{n}\sum_{k=0}^{n-1}\,\gamma^k(\obs)$ with respect to $\omega$. Then the sequence
$\bz\mu_{n,\obs}\jz_{n\in\N}$ satisfies the large deviation principle with the good rate function
\begin{equation*}
I(x)=\sup_{t\in\R}\,\{ tx-\log r(t)\}\,,
\end{equation*}
where $r(t)$ is the spectral radius of $\Et$.
\end{thm}
\begin{proof} 
By Proposition \ref{ergodicity prop}, $\Eu$ is irreducible. 
If $b$ is a non-negative element in $\B$ then $e^{ta}\otimes b\ge e^{-|t|\,\norm{a}}\unit\otimes b$ (where $\unit=\unit_{\A}$ is the unit of $\A$), hence
\begin{equation*}
\Ee_{e^{ta}}(b)=\Ee\bz e^{ta}\otimes b\jz\ge e^{-|t|\,\norm{a}}\,\Ee\bz \unit\otimes b\jz=e^{-|t|\,\norm{a}}\,\Ee_{\unit}(b)\,
\end{equation*}
i.e. $\Et\ge e^{-|t|\,\norm{a}}\,\Eu$. This implies that  
$\Et$ is also irreducible for all $t\in\R$.
  The moment generating function (\ref{momgen0}) of the G\"artner-Ellis theorem in this case is 
\begin{equation*}
\mg_n(t)=\omega_n\bz \bz e^{ta}\jz^{\otimes n}\jz=\rho\bz \Ee_{e^{ta}}^n(\unit_{\B})\jz
\end{equation*}
and by Lemma \ref{convergence}
\begin{equation*} 
\frac{1}{n}\,\log\,\mg_n(t)\xrightarrow[n\to\infty]{} \log r(t)\, ,
\end{equation*}
where $r(t)$ is the spectral radius of $\Et$. Since $t\mapsto\Et$ is analytic and $r(t)$ is a simple eigenvalue due to irreducibility, the function $t\mapsto r(t)$ is $C^{\infty}$, a standard fact in perturbation theory \cite{Rellich,Kato}. The G\"artner-Ellis theorem then yields the desired statement.
\end{proof}

\section{Factorization of states}\label{factorization section}

\begin{defin}
A translation-invariant state $\omega$ on a spin chain satisfies the \ki{upper (lower) factorization property} if there exists a constant $\beta\in\R$ ($\alpha\in\R$) and an $m_0\in\N$ such that for all $m\ge m_0$ and $k\in\N$ we have 
 \begin{eqnarray}
 \omega_{[1,km]}&\le& \beta^{k-1}\,\omega_{[1,m]}^{\otimes k}\ds\ds\ds\text{(upper factorization)}\label{upperfact2}\\
 \omega_{[1,km]}&\ge& \alpha^{k-1}\,\omega_{[1,m]}^{\otimes k} \ds\ds\ds\text{(lower factorization)}.\label{lowerfact2}
 \end{eqnarray}
\end{defin}

Note that the choice $k=2$ in the factorization properties imply  
\begin{eqnarray}
\omega&\le& \beta\,\omega_{(-\infty,0]}\otimes\omega_{[1,+\infty)} \label{upperfact3}\\
\omega&\ge& \alpha\,\omega_{(-\infty,0]}\otimes\omega_{[1,+\infty)} \label{lowerfact3}
\end{eqnarray}
due to the translation-invariance of $\omega$. On the other hand, \eqref{upperfact3} and \eqref{lowerfact3} imply \eqref{upperfact2} and \eqref{lowerfact2}, respectively, with any choice of $m$ and $k$. Hence $m_0$ doesn't actually play a role in the factorization properties, and it is enough to check \eqref{upperfact2} and \eqref{lowerfact2} for $k=2$.

Obviously, product states satisfy both lower and upper factorization. As non-trivial examples, we consider Gibbs states of translation-invariant finite-range interactions and finitely correlated states. 

\begin{lemma}\label{Gibbsfact}
Let $\vfi$ be the (unique) Gibbs state of a translation-invariant finite-range interaction $\inta$. Then $\vfi$ satisfies both lower and upper factorization properties. 
\end{lemma}
\begin{proof}
We use the notations and results of Appendix \ref{gibbs}. Let  
\begin{equation*}
W_0:=\sum \big\{\inta(X)\,:\,X\cap [-\di(\inta),0]\ne\emptyset,\,X\cap [1,\di(\inta)]\ne\emptyset \big\}
\end{equation*}
be the interaction term through site $0$. For $m\ge 2\di(\inta)$  we
define the perturbation operator
\begin{equation*}
\Q:=-\bz W_0+\gamma^m(W_0)+\gamma^{2m}(W_0)\jz\,,
\end{equation*}
which is the negative interaction term through the sites $0,m,2m$.
Then for $n>2m$ the perturbed Hamiltonian is
\begin{equation*}
H_{[-n,n]}+\Q=H_{[-n,0]}+H_{[1,m]}+H_{[m+1,2m]} +H_{[2m+1,n]} \,,
\end{equation*}
a sum of commuting terms, hence the perturbed local Gibbs state is
\begin{equation*}
\vfi^{\Q}_{[-n,n],G}=\vfi_{[-n,0],G}\otimes\vfi_{m,G}\otimes\vfi_{m,G}\otimes \vfi_{[2m+1,n],G}\,.
\end{equation*}
Since the thermodynamical limit of the above sequence of states gives the global perturbed state $\vfi^{\Q}$, we obtain 
\begin{equation*}
\vfi^{\Q}_{[1,2m]}=\vfi_{m,G}\otimes\vfi_{m,G}\,.
\end{equation*}
We now apply Corollary \ref{LRB3} with the choice $Q:=\Q$ and $L:=2\di(\inta)$. Then both $\norm{Q}_{\d\le L}$ and $\norm{Q}$ are upper bounded by $3\norm{W_0}$ and
\begin{equation*}
\vfi^{\Q}(a)\,e^{-3(1+c_L)\norm{W_0}}\le\vfi(a)\le \vfi^{\Q}(a)\,e^{3(1+c_L)\norm{W_0}}\,
\end{equation*}
for any $a\in\algloc$. This results in
\begin{equation}\label{gibbsfact}
 e^{-3(1+c_L)\norm{W_0}}\,\vfi_{m,G}\otimes\vfi_{m,G} 
\le\vfi_{[1,2m]}
\le e^{3(1+c_L)\norm{W_0}}\,\vfi_{m,G}\otimes \vfi_{m,G}\,.
\end{equation}
Now taking $\lambda$ from Lemma \ref{Araki} and 
$\beta:=\lambda^2\, e^{3(1+c_L)\norm{W_0}},\s\alpha:=\frac{1}{\beta}$ we obtain
\begin{equation*}
\alpha\,\vfi_{[1,m]}\otimes\vfi_{[1,m]}\le \vfi_{[1,2m]}\le\beta\,\vfi_{[1,m]}\otimes\vfi_{[1,m]}\,.\qedhere
\end{equation*}
\end{proof}
Note that Lemma \ref{Gibbsfact} with Lemma \ref{Araki} implies that the local Gibbs states satisfy the factorization property
\begin{equation}\label{gibbsfact2}
\tilde\beta^{-k}\,\vfi_{m,G}^{\otimes k}\otimes\vfi_{[km+1,n],G} 
\le\vfi_{n,G}
\le \tilde\beta^{k}\,\vfi_{m,G}^{\otimes k}\otimes\vfi_{[km+1,n],G}\,,\ds\ds\ds km+1\le n\le (k+1)m
\end{equation}
with some $\tilde\beta>1$.
\medskip

Now we turn to the case of finitely correlated states. For a brief introduction and notations see Appendix \ref{fcs}. To prove upper factorization for finitely correlated states, the following lemma plays the key role:
\begin{Lemma}\label{factlemma}
Let $\B\subset\B(\iK)$ be a finite-dimensional $C^*$-subalgebra of $\B(\iK)$ for some finite-dimensional Hilbert space $\iK$. Let $\rho$ be a faithful state on $\B$ and 
 $\tmap:\,\B\to\B$ be the completely positive unital map $b\mapsto \rho(b)\,\unit_{\B}$.
Then there exists a constant $\beta>1$ such that $\id_{\B}\le\cp \beta\tmap$, where $\le\cp$ means the order of complete positivity. 
\end{Lemma}
 \begin{proof}
Let $\D{\rho}=\sum_{i=1}^{\dim\iK}\,r_i\,\pr{e_i}$ be an eigen-decomposition of $\D{\rho}$ and $\beta:=\dim\iK/\min\{r_i\}$. With $\psi:=\sum_i\,e_i\otimes e_i$ we have 
\begin{equation*}
\id_{\B}\otimes \bz \beta\tmap-\id_{\B}\jz\,\pr{\psi}=\beta\,\D{\rho}\otimes\unit_{\B}-\pr{\psi}\ge 0\, ,
\end{equation*}
which, by Choi's characterization of complete positivity \cite{Choi}, gives the desired statement.
\end{proof}

\begin{prop}\label{fact}
Finitely correlated states satisfy the upper factorization property. 
\end{prop}
\begin{proof}
Let $\omega$ be a finitely correlated state with generating triple $(\B,\Ee,\rho)$ as given in (\ref{fcdef1}) and (\ref{fcdef2}).
By Lemma \ref{factlemma} we have $\id_{\A}^{\otimes k}\otimes\Ee^*\le \beta \id_{\A}^{\otimes k}\otimes \bz\Ee^*\circ\tmap^*\jz$ for any $k\in\N$, hence by (\ref{fcdef1}) for any $n>k$
\begin{eqnarray*}
\D{\vfi}_n&\le&  \beta\,(\id_{\A}^{\otimes (n-1)}\otimes \Ee^*)\circ\ldots \circ(\id_{\A}^{\otimes (k+1)}\otimes \Ee^*)\circ(\id_{\A}^{\otimes k}\otimes (\Ee^*\circ\tmap^*))\\
& &\circ(\id_{\A}^{\otimes (k-1)}\otimes \Ee^*)\circ\ldots\circ(\id_{\A}\otimes \Ee^*)\circ\Ee^*\,(\D{\rho})\\
&=&\beta\,(\id_{\A}^{\otimes (n-1)}\otimes \Ee^*)\circ\ldots\circ (\id_{\A}^{\otimes (k+1)}\otimes \Ee^*)\circ(\id_{\A}^{\otimes k}\otimes (\Ee^*\circ\tmap^*))\,\D{\vfi}_k\,.
\end{eqnarray*}
Since $\tmap^*(b)=\D{\rho}\,\Tr b$, we have $\id_{\A}^{\otimes k}\otimes (\Ee^*\circ\tmap^*) x=(\Tr_{\B}x)\otimes\D{\vfi_1}$
for any $x\in\A^{\otimes k}\otimes\B$; in particular $(\id_{\A}^{\otimes k}\otimes (\Ee^*\circ\tmap^*))\,\D{\vfi}_k=\D{\omega}_k\otimes\D{\vfi}_1$. Thus
\begin{eqnarray*}
 \D{\vfi}_n&\le&  \beta\,(\id_{\A}^{\otimes (n-1)}\otimes \Ee^*)\circ\ldots \circ(\id_{\A}^{\otimes (k+1)}\otimes \Ee^*) \,\D{\omega}_k\otimes\D{\vfi}_1\\
&=& \beta\,\D{\omega}_k\otimes\left[
(\id_{\A}^{\otimes(n-k-1)}\otimes
\Ee^*)\circ\dots\circ (\id_{\A}
\otimes\Ee^*)\D{\vfi}_1\right]\\
&=& \beta\,\D{\omega}_k\otimes\D{\vfi}_{n-k}\, .
\end{eqnarray*}
  Taking partial trace over $\B$ in the above inequality yields $\D{\omega}_n\le\beta\,\D{\omega}_k\otimes\D{\omega}_{n-k}$. The choice $n=2k$ gives the upper factorization property.
\end{proof}
\medskip

Note that $\tmap$ in Lemma \ref{factlemma} is positivity increasing whereas $\id_{\B}$ is not, therefore a converse inequality $\id_{\B}\ge \alpha\tmap$ cannot hold with any positive constant $\alpha$. Of course, to guarantee lower factorization it would be sufficient to have $\Ee^*\ge\cp \alpha\bz\Ee^*\circ\tmap^*\jz$ with some positive constant $\alpha$. A similar computation to that of Proposition \ref{fact} shows that $\Ee^*\ge\cp \alpha\, (\id_{\A}\otimes\tmap^*)\circ\Ee^*$ with some $\alpha>0$ is also a sufficient condition for lower factorization. However, neither of these conditions can hold in general, simply because finitely correlated states do not satisfy lower factorization property in general. For instance, the following is easy to verify:
\begin{ex}
Let $\omega$ be a classical Markov chain on $\X^{\infty}$, $|\X|<\infty$ with transition matrix $T$ and faithful stationary distribution $r$. Then $\omega$ satisfies the lower factorization property if and only if $T>0$ (i.e.~all entries are strictly positive).
\end{ex}
A class of examples for non-classical finitely correlated states satisfying the lower factorization property is provided by the following:
\begin{ex}
Let $\omega$ be a finitely correlated state as in Example \ref{commaux} (i.e. with a commutative algebra in the generating triple). Then 
\begin{enumerate}
\item\label{lf1}
there exists $\alpha>0$ such that $\Ee^*\ge\cp \alpha\bz\Ee^*\circ\tmap^*\jz$  if and only if
\begin{equation}\label{lf11}
T>0\ds\ds\ds\text{and}\ds\ds\ds\supp\state_{xy}=\supp\state_{zy}\,;\ds x,y,z\in\X\,;
\end{equation}
\item\label{lf2}
there exists $\alpha>0$ such that  $\Ee^*\ge\cp \alpha\, (\id_{\A}\otimes\tmap^*)\circ\Ee^*$  if and only if
\begin{equation}\label{lf21}
T>0\ds\ds\ds\text{and}\ds\ds\ds\supp\state_{xy}=\supp\state_{xz}\,;\ds x,y,z\in\X\,.
\end{equation}
\end{enumerate}
\end{ex}
\begin{proof}
We only prove \ref{lf1} as the proof of \ref{lf2} is completely similar. Note that since $\Ee^*$ and $\Ee^*\circ\tmap^*$ map from a commutative algebra then complete positivity ordering coincides with the usual positivity ordering, which in turn is enough to check on each minimal projection $\D{\delta}_x$. Now
\begin{equation*}
\bz\Ee^* -\alpha\bz\Ee^*\circ\tmap^*\jz\jz\D{\delta}_x=
\sum_y\,T_{xy}\D{\state}_{xy}\otimes\D{\delta}_y-\alpha\Ee^*\D{\state}
=\sum_y\,\bz T_{xy}\D{\state}_{xy}-\alpha\,\sum_{z}\,r_z T_{zy}\D{\state}_{zy}\jz\otimes\D{\delta}_y\,,
\end{equation*}
which is positive if and only if 
\begin{equation}\label{eq1}
T_{xy}\D{\state}_{xy}-\alpha\,\sum_{z}\,r_z T_{zy}\D{\state}_{zy}\ge 0
\end{equation}
for all $y$. Obviously, the existence of a positive $\alpha$ such that (\ref{eq1}) holds for all $x,y$ is equivalent to
\begin{equation}\label{eq2}
\supp T_{xy}\D{\state}_{xy}=\supp T_{zy}\D{\state}_{zy};\ds\ds\ds x,y,z\,,
\end{equation}
which is satisfied if (\ref{lf11}) holds. On the other hand, if $T_{xy}=0$ for some $x,y$ then by (\ref{eq2}) we get $T_{zy}=0$ for the same $y$ and all $z$. However, this contradicts the strict positivity of $r$, hence (\ref{eq2}) implies (\ref{lf11}).
\end{proof}

Note that in both examples $T$ is the matrix of $\Eu$ thus the condition $T>0$ is equivalent to $\Eu$ being positivity increasing. This condition is stronger than the strong mixing property (see Appendices \ref{fcs} and \ref{pos}).

\section{Pressure and large deviation upper bound}\label{existence}
 
\begin{thm}\label{pressure theorem}
If $\omega$ satisfies the upper factorization property then the pressure
\begin{equation*}
 \pressure{\omega}{\inta}:=\lim_{n\to\infty}\,\frac{1}{n}\log\omega\bz e^{-H_{[1,n]}}\jz
\end{equation*}
exists for any translation-invariant finite-range interaction $\inta$, where $H_{[1,n]}$ are the local Hamiltonians of $\inta$. Moreover, 
\begin{enumerate}
\item\label{boundedness}
$\displaystyle{\abs{\pressure{\omega}{\inta}}\le \norm{A_{\inta}}}$, where 
$\displaystyle{A_{\inta}:=\sum_{0\in X\in\Pn}\,\frac{1}{|X|}\inta(X)}$
is the mean energy of the interaction $\inta$;
\item\label{convexity}
the function $\momgen{\omega}{\inta}(t):=\pressure{\omega}{t\inta}$ on $\R$ is convex and Lipschitz-continuous with Lipschitz constant $\norm{A_{\inta}}$.
 \end{enumerate}
\end{thm}
\begin{proof}
Let $\vfi$ be the global Gibbs state of $\inta$ and $\bbeta$ be the upper factorization constant for $\omega$.
Inequality (\ref{gibbsfact2}) implies that 
\begin{equation*}
 e^{-H_{[1,n]}}\le \tilde\beta^{k}\,\frac{\Tr e^{-H_{[1,n]}}}{\bz\Tr e^{-H_{[1,m]}}\jz^k}\bz\, e^{-H_{[1,m]}}\jz^{\otimes k}\otimes I_{[km+1,n]}
 \end{equation*}
for any $km+1\le n \le(k+1)m$. Then with $H_n:=H_{[1,n]}$ we have
\begin{equation}\label{upperinequ}
\frac{1}{n}\log\omega\bz e^{-H_n}\jz\le
 \frac{1}{n}\log\Tr e^{-H_{n}}-\frac{k}{n}\log\Tr e^{-H_m}+\frac{k}{n}\log\tilde\beta+\frac{1}{n}\log\omega_{km}\bz \bz e^{-H_m}\jz^{\otimes k} \jz.
\end{equation}
By the upper factorization property of $\omega$,
\begin{equation*}
\omega_{km}\bz \bz e^{-H_m}\jz^{\otimes k} \jz \le \bbeta^{k-1}\omega_m^{\otimes k}\bz \bz e^{-H_m}\jz^{\otimes k}\jz=
\bbeta^{k-1}\bz\omega_m\bz e^{-H_m}\jz\jz^k\,,
\end{equation*}
hence by \eqref{upperinequ}
\begin{equation*}
\limsup_{n\to\infty}\, \frac{1}{n}\log\omega\bz e^{-H_n}\jz
\le P(\inta)-\frac{1}{m}\log \Tr e^{-H_m}+\frac{1}{m}\log (\bbeta\tilde\beta)+\frac{1}{m}\log \omega_{m}\bz e^{-H_m}\jz\,,
\end{equation*}
where $P(\inta):=\lim_{n\to\infty}\,\frac{1}{n}\log\Tr e^{-H_n}$ is the pressure of $\inta$. 
Now we can take the liminf in $m$ and obtain
\begin{equation*}
\limsup_{n\to\infty}\,\frac{1}{n}\log\omega\bz e^{-H_n}\jz\le \liminf_{n\to\infty}\,\frac{1}{n}\log \omega\bz e^{-H_n}\jz\,.
\end{equation*}

Property \ref{boundedness} is obvious from $\abs{\frac{1}{n}\log\omega\bz e^{-H_n}\jz}\le\frac{1}{n}\norm{H_n}$ and $ \limsup\nolimits_n\frac{1}{n}\norm{H_n}\le \norm{A_{\inta}}$, where the latter follows from
\begin{equation*}
\norm{\frac{1}{n}H_n-\frac{1}{n}\sum_{k=1}^n\,\gamma^k\bz A_{\inta}\jz}\le \frac{1}{n}\sum_{k=1}^n\sum\left\{
\frac{\norm{\inta(X)}}{|X|}\,:\,X\ni k,\s X\cap\bz\Z\setminus [1,n]\jz\ne\emptyset
\right\}\xrightarrow[n\to\infty]{} 0\,.
\end{equation*}

 Straightforward computation shows that the second derivative of $t\mapsto\log\omega\bz e^{tH_{n}}\jz$ is equal to the variance of the probability measure
\begin{equation*}
\mu_t(B):=\omega\bz \egy_B(H_{n})e^{tH_{n}}\jz/\omega\bz e^{tH_{n}}\jz\,;\ds\ds\ds B\subset\R\,,
\end{equation*}
therefore 
$t\mapsto \frac{1}{n}\log\omega\bz e^{tH_{n}}\jz$ is convex for all $n$, which implies the convexity of $\momgen{\omega}{\inta}$. Lipschitz continuity is easily verified from
$e^{-\abs{s-t}\norm{H_n}}e^{tH_n}\le e^{sH_n}\le e^{\abs{s-t}\norm{H_n}}e^{tH_n}\,;\s t,s\in\R$.
\end{proof}
\begin{cor}\label{ld upper bound}
Let $\inta$ be a translation-invariant finite-range interaction with local Hamiltonians $H_{[1,n]}$ and let 
\begin{equation*}
\mu_{n,\inta}(B):=\omega\bz\egy_B\bz\frac{1}{n}H_{[1,n]}\jz \jz\,;\ds\ds\ds B\subset\R\,
\end{equation*}
be the distribution of $\frac{1}{n}H_{[1,n]}$ with respect to the state $\omega$. If $\omega$ satisfies the upper factorization property then the sequence $\bz\mu_{n,\inta}\jz_{n\in\N}$ satisfies the large deviation upper bound with the rate function
\begin{equation*}
I(x)=\sup_{t\in\R}\{ tx-\pressure{\omega}{-t\inta}\}\,.
\end{equation*}
\end{cor}
\begin{proof}
It is immediate from the G\"artner-Ellis theorem and Theorem \ref{pressure theorem}.
\end{proof}

\begin{remark}
Let $\omega$ be a completely ergodic finitely correlated state. As we have seen in the proof of Theorem \ref{fcldp}, 
\begin{equation*}
\lim_{k\to\infty}\,\frac{1}{k}\,\log\,\omega_{km}\bz \bz e^{-H_m}\jz^{\otimes k}\jz=\log r\bz \Ee_{e^{-H_m}}\jz\,,
\end{equation*}
hence \eqref{upperinequ} yields
\begin{equation*}
\limsup_{n\to\infty}\, \frac{1}{n}\log\omega\bz e^{-H_n}\jz
\le P(\inta)-\frac{1}{m}\log \Tr e^{-H_m}+\frac{1}{m}\log (\bbeta\tilde\beta)+\frac{1}{m}\log r\bz \Ee_{e^{-H_m}}\jz\,,
\end{equation*}
and by taking the liminf in $m$ we get
\begin{equation*}
\limsup_{n\to\infty}\, \frac{1}{n}\log\omega\bz e^{-H_n}\jz
\le \liminf_{m\to\infty}\frac{1}{m}\log r\bz \Ee_{e^{-H_m}}\jz\,.
\end{equation*}
Using the lower factorization property for the local Gibbs states, a similar argument gives
\begin{equation*}
\liminf_{n\to\infty}\, \frac{1}{n}\log\omega\bz e^{-H_n}\jz
\ge \limsup_{m\to\infty}\frac{1}{m}\log r\bz \Ee_{e^{-H_m}}\jz\,,
\end{equation*}
hence 
\begin{equation*}
\pressure{\omega}{\inta}=\lim_{n\to\infty}\,\frac{1}{n}\log\omega\bz e^{-H_{n}}\jz=\lim _{n\to\infty}\frac{1}{n}\log r\bz \Ee_{e^{-H_n}}\jz\,.
\end{equation*}
\end{remark}
 
\section{Chernoff bound}\label{chernoff section}

Assume that a sequence of finite-level systems with Hilbert spaces $\sequence{\iH}:=\{\iH_n\,:\,n\in\N\}$ is given together with
 two sequences of states $\sequence{\omega}:=\{\omega_n\,:\,n\in\N\}$ and $\sequence{\sigma}:=\{\sigma_n\,:\,n\in\N\}$. The true states of the systems are unknown, but we know a priori that with probability $0<\kappa<1$ the systems are in state $\omega_n$ for all $n\in\N$ (null-hypothesis $H_0$) and with probability $1-\kappa$ the systems are in state $\sigma_n$ for all $n\in\N$ (counter-hypothesis $H_1$).
To decide between these two options we make a binary measurement on a system for some $n$ with measurement operators $0\le A_n\le I_{\iH_n}$ corresponding to outcome $0$ and $I_{\iH_n}-A_n$  corresponding to outcome $1$;
if the outcome is $0$ (resp.~$1$) then hypothesis $H_0$ (resp.~$H_1$) is accepted.
The Bayesian probability of an erroneous decision is
\begin{equation*} 
\pn{\omega_n}{\sigma_n}:=\kappa\,\omega_n(I-A_n)+(1-\kappa)\,\sigma_n(A_n)\,.
\end{equation*}
Let
\begin{equation}\label{minimum error}
\pmin{\omega_n}{\sigma_n}:=\min\{\pn{\omega_n}{\sigma_n}\,:\,0\le A_n\le I\}\,.
\end{equation}
It is easy to see that the best test achieving (\ref{minimum error}) is the spectral projection of $\kappa\D{\omega}_n-(1-\kappa)\D{\sigma}_n$ corresponding to the positive part of the spectrum and therefore
\begin{equation*}
\pmin{\omega_n}{\sigma_n}=\frac{1}{2}-\half\norm{\kappa\D{\omega}_n-(1-\kappa)\D{\sigma}_n}_1\,.
\end{equation*}
We are interested in the asymptotics of $\frac{1}{n}\log\pmin{\omega_n}{\sigma_n}$ for large $n$'s. Obviously the value of $\kappa$ doesn't play a role here, hence we may just as well assume $\kappa=\half$.
 
It was shown in \cite{Aud} that
\begin{equation*}
\pmin{\omega_n}{\sigma_n}\le\half\Tr\D{\omega}_n^{1-t}\D{\sigma}_n^{t}
\end{equation*}
for all $0\le t\le 1$. 
(Here $A^0$ is defined to be the support projection of $A$ for a positive semidefinite $A\in\B(\iH)$, so that 
$t\mapsto \Tr\D{\omega}_n^{1-t}\D{\sigma}_n^{t}$ is continuous on $[0,1]$.)
Now if 
\begin{equation}\label{chernoff bound}
\chb{t}{\sequence{\omega}}{\sequence{\sigma}}:=\lim_{n\to\infty}\,\frac{1}{n}\,\log\Tr\D{\omega}_n^{1-t}\D{\sigma}_n^{t}
\end{equation}
 exists for all $0\le t\le 1$ then we get
\begin{equation}\label{upper}
\limsup_{n\to\infty}\frac{1}{n}\log\pmin{\omega_n}{\sigma_n}\le\inf_{0\le t\le 1}\chb{t}{\sequence{\omega}}{\sequence{\sigma}}\,.
\end{equation}
The quantity on the right-hand side of (\ref{upper}) is called the \ki{Chernoff bound}.
In the case $\iH_n=\iH_1^{\otimes n};\ds \omega_n=\omega_1^{\otimes n},\ds \sigma_n=\sigma_1^{\otimes n}$ the limit (\ref{chernoff bound}) obviously exists and hence (\ref{upper}) holds with $\chb{t}{\sequence{\omega}}{\sequence{\sigma}}=\min_{0\le t\le 1}\Tr \D{\omega}_1^{1-t}\D{\sigma}_1^{t}$. The result of \cite{SzN} shows that in this case also 
\begin{equation*}
\liminf_{n\to\infty}\frac{1}{n}\log\pmin{\omega_n}{\sigma_n}\ge\min_{0\le t\le 1}\chb{t}{\sequence{\omega}}{\sequence{\sigma}}\,.
\end{equation*}

One can consider the above situation as discriminating between two product states on the infinite spin chain by making measurements on finite parts of the chain. We show that the above results can be extended to a wider class of states with suitable factorization properties:
\begin{thm}\label{chernoff theorem}
Let $\omega$ and $\sigma$ be translation-invariant states on a spin chain.
\begin{enumerate}
\item\label{chernoff theorem upper}
If $\omega$ and $\sigma$ satisfy the upper factorization property then
for all $0\le t\le 1$ the limit 
\begin{equation}\label{chb1}
\chb{t}{\omega}{\sigma}:=\lim_{n\to\infty}\,\frac{1}{n}\log\Tr\D{\omega}_n^{1-t}\D{\sigma}_n^t\,
\end{equation}
exists and 
\begin{equation*}
\limsup_{n\to\infty}\,\frac{1}{n}\log\pmin{\omega_n}{\sigma_n}\le \inf_{0\le t\le 1}\,\chb{t}{\omega}{\sigma}\,.
\end{equation*}
\item\label{chernoff theorem lower}
If $\omega$ and $\sigma$ satisfy both upper and lower factorization properties then
$t\mapsto \chb{t}{\omega}{\sigma}$ is continuous on $[0,1]$, and
\begin{equation}\label{chernoff limit}
\lim_{n\to\infty}\,\frac{1}{n}\log\pmin{\omega_n}{\sigma_n}= \min_{0\le t\le 1}\,\chb{t}{\omega}{\sigma}\,.
\end{equation}
\end{enumerate}
\end{thm}
\begin{proof}
(i)\ds
 We can assume that $\omega$ and $\sigma$ have the same upper factorization constant $\beta$. Let $n>m\ge 1$ and write $n=km+r$ with $1\le r\le m$. By the operator monotonicity of the function $x\mapsto x^t,\s x\in [0,\infty)$ \cite{Bhatia} we get 
\begin{equation*}
\Tr\D{\omega}_n^{1-t}\D{\sigma}_n^t\le \beta^{k}\,M\,\bz\Tr\D{\omega}_m^{1-t}\D{\sigma}_m^t\jz^k
\end{equation*}
with $M:=\max\{\Tr\D{\omega}_{r}^{1-t}\D{\sigma}_r^t \,:\,1\le r\le m\}$, hence
\begin{equation}\label{upper2}
\limsup_{n\to\infty}\,\frac{1}{n}\log\Tr\D{\omega}_n^{1-t}\D{\sigma}_n^t\le \frac{1}{m}\log\beta+\frac{1}{m}\log\Tr\D{\omega}_m^{1-t}\D{\sigma}_m^t
\end{equation}
for all $m$. Taking the liminf in $m$ gives the existence of the limit and \eqref{upper} gives the rest of the statement.

(ii)\ds
 We can also assume that $\omega$ and $\sigma$ have the same lower factorization constant $\alpha$.
In the same way as above, lower factorization property implies
\begin{equation}
\frac{1}{m}\log\alpha+\frac{1}{m}\log\Tr\D{\omega}_m^{1-t}\D{\sigma}_m^t\le\liminf_{n\to\infty}\,\frac{1}{n}\log\Tr\D{\omega}_n^{1-t}\D{\sigma}_n^t\,.
\end{equation}
Combining it with \eqref{upper2} we get that $t\mapsto\chb{t}{\omega}{\sigma}$ is the uniform limit of the continuous functions 
$t\mapsto\frac{1}{m}\log\Tr\D{\omega}_m^{1-t}\D{\sigma}_m^t$, hence continuity of the limit follows.

 Let $A_n$ be the optimal test for discriminating between $\omega_n$ and $\sigma_n$ and $n=km+r$ as before. Then
\begin{eqnarray*}
\pmin{\omega_n}{\sigma_n}&=&
\pn{\omega_n}{\sigma_n}=\pn{\omega_{(k+1)m}}{\sigma_{(k+1)m}}\\
&\ge& \alpha^{k}\,\pn{\omega_m^{\otimes(k+1)}}{\sigma_m^{\otimes(k+1)}}\\
&\ge& \alpha^{k}\,\pmin{\omega_m^{\otimes(k+1)}}{\sigma_m^{\otimes(k+1)}}
\end{eqnarray*}
hence
\begin{equation*}
\liminf_{n\to\infty}\,\frac{1}{n}\log\pmin{\omega_n}{\sigma_n}\ge \frac{1}{m}\log \alpha+\frac{1}{m}
\lim_k\,\frac{1}{k}\,\log\pmin{\omega_m^{\otimes(k+1)}}{\sigma_m^{\otimes(k+1)}}\,,
\end{equation*}
where the last limit is known to exist due to \cite{SzN,Aud} where it was also shown to be equal to
$\min_{0\le t\le 1}\,\log\Tr\D{\omega}_m^{1-t}\D{\sigma}_m^{t}$. Now by (\ref{upper2}) we have
\begin{equation*}
\frac{1}{m}\min_{0\le t\le 1}\,\log\Tr\D{\omega}_m^{1-t}\D{\sigma}_m^{t}\ge \min_{0\le t\le 1}\,\chb{t}{\omega}{\sigma}-\frac{1}{m}\log\beta\,,
\end{equation*}
thus 
\begin{equation*}
\liminf_{n\to\infty}\,\frac{1}{n}\log\pmin{\omega_n}{\sigma_n}\ge  \min_{0\le t\le 1}\,\chb{t}{\omega}{\sigma}-\frac{1}{m}\log\frac{\beta}{\alpha}\,.
\end{equation*}
Since this holds for all $m$ we get
\begin{equation*}
\liminf_{n\to\infty}\,\frac{1}{n}\log\pmin{\omega_n}{\sigma_n}\ge  \min_{0\le t\le 1}\,\chb{t}{\omega}{\sigma}.
\end{equation*}
\end{proof}

\begin{ex}
Let $\omega$ and $\sigma$ be completely ergodic finitely correlated states with commutative auxiliary algebras $\F(\X)$ and $\F(\Y)$, respectively, with local restrictions
\begin{eqnarray*}
  \D{\omega}_n&=&\sum_{x_1,\ldots, x_{n}\in\X}\,r_{x_1} \bz T_{x_1x_2}\D{\state}_{x_1\,x_2}\jz \otimes\ldots\otimes \bz T_{x_{n-1}x_n}\D{\state}_{x_{n-1}\,x_{n}}\jz\otimes\D{\Theta}_{x_n}\\
  \D{\sigma}_n&=&\sum_{y_1,\ldots, y_{n}\in\Y}\,p_{y_1} \bz S_{y_1y_2}\D{\vfi}_{y_1\,y_2}\jz \otimes\ldots\otimes \bz S_{y_{n-1}y_n}\D{\vfi}_{y_{n-1}\,y_{n}}\jz\otimes\D{\Phi}_{y_n}\,,
\end{eqnarray*}
where $T$ and $S$ are primitive stochastic matrices with stationary distributions $\{r_x\}$ and $\{p_y\}$, respectively, and
$\D{\Theta}_{x}=\sum_{w} T_{xw}\D{\state}_{xw}$, $\D{\Phi}_{y}=\sum_{z} T_{yz}\D{\state}_{yz}$ (see Example \ref{commaux} for details.)  Assume that $\supp\D{\state}_{xw}\perp\supp\D{\state}_{uv}$ whenever the first indices are different, and similarly for the states $\{\vfi_{yz}\}$. Then also $\supp\D{\Theta}_x\perp\supp\D{\Theta}_u,\s \supp\D{\Phi}_y\perp\supp\D{\Phi}_z$ when $x\ne u,\s y\ne z$, and 
\begin{eqnarray*}
\Tr\D{\omega}_n^{1-t}\,\D{\sigma}_n^t&=&\sum_{\substack{x_1,\ldots,x_{n}\\y_1,\ldots,y_{n}}}
r_{x_1}^{1-t}p_{y_1}^t\,\bz\Tr\D{\Theta}_{x_n}^{1-t}\D{\Phi}_{y_n}^t \jz\prod_{k=1}^{n-1}\,\bz T_{x_k x_{k+1}}^{1-t}S_{y_k y_{k+1}}^{t}\Tr\D{\state}_{x_k x_{k+1}}^{1-t}\D{\vfi}_{y_k y_{k+1}}^t \jz\\
&=&
\inner{a(t)}{Q(t)^{n-1}b(t)}
\end{eqnarray*}
for every $t\in\R$ with 
\begin{equation*} 
a(t)_{(x,y)}:=r_{x}^{1-t}p_{y}^t,\ds\ds b(t)_{(x,y)}:=\Tr\D{\Theta}_{x}^{1-t}\D{\Phi}_{y}^t\ds\ds\text{and}\ds\ds 
Q(t)_{(x,y),(w,z)}:=T_{x w}^{1-t}S_{y z}^{t}\Tr\D{\state}_{x w}^{1-t}\D{\vfi}_{y z}^t\,.
\end{equation*}
(Note that the convention $0^t:=0,\s t\in\R$ is used here, i.e.~$A^t$ is meant to be taken on the support of $A$ for a positive semidefinite $A$.) Now if $\Tr\D{\state}_{x w}^{1-t}\D{\vfi}_{y z}^t >0$ for all $x,y,w,z$ for some (and hence for all) $t\in\R$ then it is easy to see that $Q(t)$ is the matrix of a primitive positive map on $\F(\X)\otimes\F(\Y)$ and $a(t)$ and $b(t)$ are strictly positive vectors for all $t\in\R$, hence by Lemma \ref{convergence} we get
\begin{equation*}
\chb{t}{\omega}{\sigma}=\lim_{n\to\infty}\frac{1}{n}\log\Tr\D{\omega}_n^{1-t}\,\D{\sigma}_n^t=\log r(t)\,,
\end{equation*}
where $r(t)$ is the spectral radius of $Q(t)$. 
\end{ex}
\smallskip

The above construction provides examples for correlated quantum states with non-commuting densities for which an explicit expression is available for the asymptotic quantity $\chb{t}{\omega}{\sigma}$, and thus the Chernoff bound
$\min_{0\le t\le 1}\chb{t}{\omega}{\sigma}$ can in principle be evaluated. We can also impose condition \eqref{lf21} on the states $\omega$ and $\sigma$ above to ensure that \eqref{chernoff limit} holds. Note, however, that the lower factorization condition may be replaced by a possibly more natural condition here. Indeed, 
$t\mapsto \chb{t}{\omega}{\sigma}=\log r(t)$ turns out to be the logarithmic moment generating function of the sequence of random variables $Z_n:=\frac{1}{n}\log\frac{q_n}{p_n}$, where $p_n$ and $q_n$ are the classical probability measures assigned to $\omega_n$ and $\sigma_n$ by the method of \cite{SzN}. As $r(t)$ is a simple eigenvalue of $Q(t)$, the function $t\mapsto\log r(t)$ is differentiable (see e.g.~\cite{Kato}), and an application of the G\"artner-Ellis theorem yields \eqref{chernoff limit}. For a more detailed argument we refer to \cite{HMO}.

 Note that in the above construction both $\omega$ and $\sigma$ contain only classical correlations between  different sites of the spin chain (i.e.~$\omega_n$ and $\sigma_n$ are both separable for all $n\in\N$). However, one can easily modify the above construction to obtain states with non-classical correlations by considering a spin chain $\C$ whose one-site algebra is split in the form $\A=\A_{\mathrm{L}}\otimes\A_{\mathrm{R}}$.
Now if we do the above construction on the spin chain $\tilde\C$ with one-site algebra $\tilde\A=\A_{\mathrm{R}}\otimes\A_{\mathrm{L}}$ then the resulting states can be considered as states on $\C$ and can contain non-classical correlations.
 
It seems to be rather non-trivial to give an explicit expression for the Chernoff bound when both $\omega$ and $\sigma$ are Gibbs states.
However, a lower bound is easy to give in terms of pressures.  Indeed, let $\inta$ and $\Psi$ denote the finite-range translation-invariant interactions for which $\omega$ and $\sigma$ are the unique Gibbs states, and let $H_n(\inta)$ and $H_n(\Psi)$ denote the corresponding local Hamiltonians. Then by Lemma \ref{Araki} and the Golden-Thompson inequality we have
\begin{eqnarray*}
\chb{t}{\omega}{\sigma}&=&\lim_{n\to\infty}\frac{1}{n}\log\Tr\omega_{n,G}^{1-t}\sigma_{n,G}^{t}\\
&=&\lim_{n\to\infty}\frac{1}{n}\left[\log\Tr e^{-(1-t)H_n(\inta)} e^{-tH_n(\Psi)}-(1-t)\log\Tr e^{-H_n(\inta)}
-t\log\Tr e^{-H_n(\Psi)}\right]\\
&\ge&\lim_{n\to\infty}\frac{1}{n}\left[\log\Tr e^{-(1-t)H_n(\inta)-tH_n(\Psi)}-(1-t)\log\Tr e^{-H_n(\inta)}
-t\log\Tr e^{-H_n(\Psi)}\right]\\
&=& P\bz (1-t)\inta+t\Psi\jz-(1-t)P(\inta)-tP(\Psi)\,.
\end{eqnarray*}
Note that this lower bound is not sharp in general, as one can easily see in the case when both $\omega$ and $\sigma$ are product states.

\section{Concluding remarks}\label{concluding remarks}

As it is usual when extending classical results to the quantum setting, large deviation questions on a spin chain may have several different formulations. In \cite{Bjelakovic} a quantum version of Sanov's theorem was presented under a certain factorization property called $^*$-mixing. It is easy to see that those of our results that rely only on the upper factorization property (Theorem \ref{pressure theorem}, Corollary \ref{ld upper bound} and \ref{chernoff theorem upper} of Theorem \ref{chernoff theorem}) still hold true if only the following common weakening of our upper factorization property and the upper bound of $^*$-mixing is assumed:
\begin{defin}
A translation-invariant state $\omega$ on a spin chain satisfies the weak upper factorization property if there exist constants $0<\beta\in\R$ and $l,m_0\in\N$ such that for all $m\ge m_0$
\begin{equation*}
\omega_{\cup_{j=0}^{k-1}[j(m+l)+1,j(m+l)+m]}\le \beta^{k-1}\,\otimes_{j=0}^{k-1}\omega_{[j(m+l)+1,j(m+l)+m]}\,.
\end{equation*}
\end{defin}

\noindent In proving Theorem \ref{pressure theorem} from this condition one has to use the factorization property 
\begin{equation*}
\vfi_{n,G}\le \tilde\beta^k\,\bz\vfi_{m,G}\otimes \Tr_{[m+1,m+l]}\jz^{\otimes k}\otimes \Tr_{[k(m+l)+1,n]}\,,\ds\ds\ds k(m+l)+1\le n\le (k+1)(m+l)
\end{equation*}
that can be proven similarly to (\ref{gibbsfact2}). In the proof of the existence of the asymptotic Chernoff bound (\ref{chb1}) one has to use the general monotonicity property of quantum quasi-entropies (\cite[Theorem 1]{Petz}).
\medskip

As it was shown in \cite[Theorem 2.1]{HP}, the weak upper factorization property also ensures the existence of the mean relative entropy 
\begin{equation*}
\msr{\vfi}{\omega}:=\lim_{n\to\infty}\,\frac{1}{n}\,\sr{\vfi_n}{\omega_n}\,,
\end{equation*}
where
\begin{equation*}
\sr{\vfi_n}{\omega_n}:=\begin{cases}\Tr \D{\vfi}_n\bz \log\D{\vfi}_n-\log\D{\omega}_n\jz\,,& \text{ if }\supp\D{\vfi}_n\subset\supp\D{\omega}_n;\\
\infty\,, & \text{ otherwise}\,.
\end{cases}
\end{equation*}
 A fundamental result in statistical physics is the variational formula connecting the pressure and the mean entropy, providing a criterion for a state to be a global equilibrium state (see e.g.~\cite{Israel}). In \cite{HMOP} a different version of the pressure functional is introduced, which coincides with our definition \eqref{pressure} when the local densities of the reference state commute with the local Hamiltonians. An advantage of that definition is that a variational principle can be established between the pressure and the mean relative entropy. On the other hand, the pressure of \cite{HMOP} can only be considered as the moment generating function for a sequence of probability measures under the assumption that the BMV conjecture \cite{BMV,LS} holds true.

\section*{Acknowledgments}

Part of this work was done when M.M.~was a visiting fellow at the Mathematical Institute of the University of Wroclaw (sponsored by the EU Research Training
Network "QP-Applications", contract HPRN-CT-2002-00279), and later a Junior Research Fellow at the Schr\"odinger Institute in Vienna. M.M.~is grateful for the excellent working conditions provided by the members of the above institutes and particuarly indebted to Marek Bo\.zejko and Heide Narnhofer for stimulating discussions and for their kind hospitality. Partial funding by Grant-in-Aid for Scientific Research (B)17340043 (F.H.) and Grant-in-Aid for JSPS Fellows 18\,$\cdot$\,06916, as well as the JSPS fellowship P06916 (M.M.) are also gratefully acknowledged.

\vspace{1cm}

\textbf{\Large Appendices}

\appendix

\section{Perturbation of Gibbs states}\label{gibbs}
Let $\inta$ be a translation-invariant finite-range interaction.
For any $Q\in\algloc\sa$ with $r:=\min\{k\in\N:\,Q\in\alg_{[-k,k]}\}$ a perturbed interaction $\inta^Q$ can be defined by 
$\inta^Q(X):=\inta(X)+\delta_{X,[-r,r]}Q$, with local Hamiltonians $H_{[-n,n]}+Q$ for $n>r$.
 Obviously, $\inta^Q$ is also a finite-range interaction with bounded surface energy, hence
the perturbed dynamics $\aut^Q$ and the perturbed Gibbs state $\vfi^Q$
corresponding to $\inta^Q(X)$
are given by the thermodynamic limits 
\begin{equation*}
\aut_t^Q(a)=\lim_{n\to\infty}\,e^{it(H_{[-n,n]}+Q)}\,a\,e^{-it(H_{[-n,n]}+Q)}\,,\ds\ds
\vfi^Q(a)=\lim_{n\to\infty}\,\frac{\tau\bz e^{-(H_{[-n,n]}+Q)}a\jz}{\tau\bz e^{-(H_{[-n,n]}+Q)}\jz}\,,\ds\ds\ds a\in\alg.
\end{equation*}
The following important bound was obtained in \cite{LRB}:
\begin{Lemma} \label{LRB1}
For every positive local element $a\in\algloc$
\begin{equation*}
|\log\ffi^Q(a)-\log\ffi(a)|
\le\|Q\|+\sup_{0\le t\le1}\,\sup_{-1/2\le s\le1/2}\|\alpha_{is}^{tQ}(Q)\|\,.
\end{equation*}
Note that the right-hand side does not depend on $a$.
\end{Lemma}
The following result is contained in \cite{Araki}, even though not explicitly stated. For readers' convenience we give a very brief argument here to show how the present form can be obtained, and refer to \cite[\S 4]{Araki} for the main body of the proof.
\begin{Lemma} \label{LRB2}
Let $Q:=-\sum_{k=1}^m \inta(\Lambda_k)$ for different finite subsets $\Lambda_1,\ldots,\Lambda_m$ of $\Z$.
For $L\ge \di(\Phi)$ let
\begin{equation*}
F_L(x):=\exp\Biggl((L-\d(\Phi)+1)x+2\sum_{j=1}^{\d(\Phi)}\frac{e^{jx-1}}{j}\Biggr)
\end{equation*}
and
\begin{equation*}
\|Q\|_{\d\le L}:=\inf\Biggl\{\sum_i\|a_i\|:Q=\sum_ia_i,\,\di(a_i)\le L\Biggr\}
\end{equation*}
where $\di(a):=\min\{\diam(\Lambda):a\in\alg\loc\}$. Then
\begin{equation*}
\|\alpha_z^{tQ}(Q)\|\le\|Q\|_{\d\le L}F_L(2|z|\,\|\Phi\|')
\quad\mbox{for all $z\in\bC$}, \s 0\le t\le 1\,,
\end{equation*}
with $\|\Phi\|':=\sum_{\Lambda\subset[0,\d(\Phi)]}\|\Phi(\Lambda)\|$.
\end{Lemma}
\begin{proof}
For $a\in\C$ let $\delta_a$ denote the bounded derivation $b\mapsto i(ab-ba)$, and 
let
\begin{equation*}
\tilde\Phi(X):=\begin{cases}
\Phi(X) & \text{if $X\ne\Lambda_1,\dots,\Lambda_k$}, \\
(1-t)\Phi(\Lambda_j) & \text{if $X=\Lambda_j$, $1\le j\le k$}.
\end{cases}
\end{equation*}
for $0\le t\le 1$. Then $\inta^{tQ}$ and $\tilde\inta$ yield the same global automorphism group $\alpha^{tQ}$ and global Gibbs state $\vfi^{tQ}$, and the generator of the former is the closure of the derivation
\begin{equation*}
\delta:=\lim_{n\to\infty}\,i\sum_{X\cap[-n,n]\ne\emptyset}\delta_{\tilde\inta(X)}\ds\ds\text{with domain}\ds\ds\mathcal{D}(\delta)=\algloc\,.
\end{equation*}
Using that 
$d(\tilde\Phi)\le d(\Phi)$ and $\|\tilde\Phi\|'\le\|\Phi\|'$, and following the arguments in \cite[\S 4]{Araki} (see also \cite[\S 9]{Araki2}) we obtain
$$
\sum_{n=0}^\infty{|z|^n\over n!}\sum_{X_1,\dots,X_n\subset\Z}
\|\delta_{\tilde\Phi(X_n)}\cdots\delta_{\tilde\Phi(X_1)}a\|
\le\|a\|_{d\le L}F_L(2|z|\,\|\Phi\|'),\qquad z\in\bC
$$
for every $a\in\algloc$. This implies that
every $a\in\algloc$ is an analytic element for $\alpha^{t Q}$ and
$$
\alpha_z^{t Q}(a)
=\sum_{n=0}^\infty{z^n\over n!}\sum_{X_1,\dots,X_n\subset\Z}
\delta_{\tilde\Phi(X_n)}\cdots\delta_{\tilde\Phi(X_1)}a,
\qquad z\in\bC,
$$
so that
$$
\|\alpha_z^{t Q}(a)\|\le\|a\|_{d\le L}F_L(2|z|\,\|\Phi\|'),
\qquad z\in\bC.
$$
Letting $a=Q$ gives the desired statement.
\end{proof}
Combining Lemmas \ref{LRB1} and \ref{LRB2} we get 
\begin{cor}\label{LRB3}
For every positive local element $a\in\algloc$
\begin{equation*}
|\log\ffi^Q(a)-\log\ffi(a)|
\le\|Q\|+\norm{Q}_{\d\le L}\,c_{L}\, ,
\end{equation*}
where the positive constant
\begin{equation*}
c_L:=\exp\Biggl((L-\d(\Phi)+1)\norm{\inta}'+2\sum_{j=1}^{\d(\Phi)}\frac{e^{j\norm{\inta}'-1}}{j}\Biggr)
\end{equation*}
depends only on the interaction $\inta$ and the choice of $L$.
\end{cor}
 
\section{Finitely correlated states}\label{fcs}

 Let $\C$ be the infinite spin chain with one-site algebra $\A\subset\B(\iH)$. The following recursive procedure to construct states on $\C$, together with a detailed analysis of the properties of the so obtained states was developed in \cite{FNW}, where states obtained by this procedure were called $C^*$-finitely correlated states, and we will refer to them simply as \ki{finitely correlated states}. Similarly to quantum Markov states \cite{AccF}, finitely correlated states provide a possible way to extend the notion of Markov chains to the quantum setting; see also \cite{Accardi} in this direction.

 For the construction one needs a triple $(\B,\Ee,\rho)$, where $\B$ is a $C^*$-subalgebra of $\B(\iK)$ for some finite dimensional Hilbert space $\iK$, $\Ee:\,\A\otimes\B\to\B$ is a unital completely positive map and $\rho$ is a faithful state on $\B$ with density operator $\D{\rho}$.
 Further, one has to assume that $\Ee$ and $\rho$ are related so that 
     \begin{equation}\label{shiftinv1}
      \Tr_{\A}\s\Ee^*\bz \D{\rho}\jz=\D{\rho}
     \end{equation}
     holds. Then 
\begin{equation}\label{fcdef1}
 \D{\vfi}_1 := \Ee^*\bz\D{\rho}\jz\, ;\ds\ds\ds\D{\vfi}_n := \bz\id_{\A}^{\otimes (n-1)}\otimes \Ee^*\jz\circ\ldots \circ\bz\id_{\A}\otimes \Ee^*\jz\circ\Ee^*\,\bz\D{\rho}\jz\, ;\ds\ds n=2,3,\ldots
\end{equation}
defines a sequence of states on $\A^{\otimes n}\otimes\B$ for each $n\in\N$.
 To obtain a family of states on the spin chain, one has to trace out the auxiliary algebra $\B$:
    \begin{equation}\label{fcdef2}
      \D{\omega}_n:=\Tr_{\B}\s\D{\vfi}_n\, .
    \end{equation}
   Compatibility of this family is guaranteed by the unitality of $\Ee$ while shift-invariance follows from 
    (\ref{shiftinv1}), hence there exists a unique translation-invariant state $\omega$ on $\C$ with local restrictions $\omega_{[1,n]}=\omega_n$. 
 
Each $a\in\A$ defines a linear map $\Ee_{a}:\,\B\to\B$ through the formula $\Ee_a (b):=\Ee(a\otimes b)$. 
   On simple product operators $\omega_n$ takes the value
     \begin{equation*}
      \omega_n(a_1\otimes\ldots \otimes a_n)=\rho\bz\Ee\bz a_1\otimes \Ee\bz a_2\otimes\ldots \Ee\bz a_n\otimes\unit_{\B}\jz\ldots \jz\jz\jz=\rho\bz \Ee_{a_1}\circ\ldots\circ\Ee_{a_n}(\unit_{\B})\jz\, .
    \end{equation*}
We use the notation $\Eu:=\Ee_{\unit_{\A}}$.
Unitality of $\Ee$ and the translation-invariant condition (\ref{shiftinv1}) can be expressed in the form
\begin{equation*} 
 \Ee_{\unit}(\unit_{\B})=\unit_{\B}\ds\ds\ds\text{and}\ds\ds\ds \Ee_{\unit}^*\bz\D{\rho}\jz=\D{\rho}\, .
\end{equation*}
  
As it was shown in \cite[Proposition 3.1]{FNW}, $\omega$ is ergodic if and only if $1$ has geometric multiplicity $1$ as an eigenvalue of $\Eu$. By Theorem \ref{PF} and Lemma \ref{PF2} we get the following equivalent version:
\begin{prop}\label{ergodicity prop}
Let $\omega$ be a finitely correlated state with generating triple $(\B,\Ee,\rho)$ with $\rho$ faithful. Then $\omega$ is ergodic if and only if $\Eu$ is irreducible.
\end{prop}
A state $\omega$ on a spin chain $\C$ can be considered as a state $\omega^{(n)}$ on the restructured chain $\C^{(n)}$ with one-site algebra $\C_{[1,n]}$. The shift of $\C^{(n)}$ is $\gamma^n$. A state $\omega$ on $\C$ is 
\ki{completely ergodic} if $\omega^{(n)}$ is ergodic for all $n\in\N$, and \ki{strongly mixing} if
\begin{equation*}
\omega\bz a\gamma^n\bz b\jz\jz\xrightarrow[n\to\infty]{} \omega(a)\omega(b)\,,\ds\ds\ds a,b\in\C\,.
\end{equation*}
Strong mixing property implies complete ergodicity in general, and 
by \cite[Proposition 1.1]{HP} for a finitely correlated state $\omega$ both properties are equivalent to $\Eu$ being a  primitive map.
\begin{ex} \ki{(commutative auxiliary algebra)}\label{commaux}
  If the auxiliary algebra $\B$ is  commutative then it is isomorphic to $\F(\X):=\{f:\,\X\to\iC\}$ for some finite set $\X$. The densities $\D{\delta}_x$ of the Dirac measures $\delta_x$ form the minimal projections of $\B$.
    $\Ee^*$ is specified by its values on $\D{\delta}_x,\s x\in\X$, and 
  since $\Ee^*\D{\delta}_x$ is a density operator
    on $\A\otimes\F(\X)$, it can uniquely be decomposed in the form 
    $\sum_y\,T_{xy}\,\D{\state}_{xy}\otimes\D{\delta}_y$, where
    $\{T_{xy}\,:\,y\in\X\}$ is a probability distribution 
    and $\state_{xy}$ are states on $\A$. The state $\rho$ has density
     $\D{\rho}=\sum_x\,r_x\,\D{\delta}_x$, and (\ref{shiftinv1}) is equivalent to $\{r_x\}$ being an invariant
    measure of the stochastic matrix $T$. The resulting finitely correlated state $\omega$ has local density operators 
    \begin{equation*}
     \D{\omega}_n=\sum_{\{x_1,\ldots, x_{n+1}\}}\,\mu_{n+1}(x_1,\ldots, x_{n+1})\,\D{\state}_{x_1\,x_2}\otimes\ldots\otimes\D{\state}_{x_n\,x_{n+1}}\, ,
    \end{equation*}
    where $\mu$ is the classical Markov measure on $\X^{\infty}$, generated by $\{r_x\}$ and $T$.
  If the states $\state_{xy}$ are independent of $y$ (or of $x$) then $\D{\omega}_n$ takes the simpler form
$\omega_n=\sum_{\{x_1,\ldots, x_n\}}\,\mu(x_1,\ldots, x_n)\,\D{\state}_{x_1}\otimes\ldots\otimes\D{\state}_{x_n}$.
  Note that the choice $\A:=\F(\X)$ and $\rho_{xy}:=\delta_x$ 
 yields the local densities $\D{\omega}_n=\sum_{\{x_1,\ldots, x_{n}\}}\,\mu_{n}(x_1,\ldots, x_{n})\,\D{\delta}_{x_1}\otimes\ldots\otimes\D{\delta}_{x_n}$, i.e.~the resulting finitely correlated state is the classical Markov measure $\mu$ on $\X^{\infty}$.
    In all these cases $\Eu$ is the linear map $T$
    on $\F(\X)$ with matrix $\{T_{xy}\,:\,x,y\in\X\}$, thus the ergodic properties of $\omega$ 
are the same as those of the classical Markov measure $\mu$.
\end{ex}

\section{Spectral properties of positive maps}\label{pos}
 
Let $\iH$ be a finite-dimensional Hilbert space and $\A\subset\B(\iH)$ be a $C^*$-subalgebra with unit $\unit$.
An element $a\in\A$ is \ki{positive} ($a\ge 0$) if it is of the form $a=x^*x$ for some $x\in\A$ and \ki{strictly positive} ($a> 0$) if there exists $\ep>0$ such that $a\ge \ep\unit$. A linear map $\tmap:\,A\to\A$  is \ki{positive} ($\tmap\ge 0$) if it maps positive elements into positive elements, and \ki{positivity increasing} ($\tmap>0$) if it maps non-zero positive elements into strictly positive elements. $\tmap$ is called \ki{unital} if $\tmap(\unit)=\unit$.
If $\tmap$ is positive then $\norm{\tmap}=\norm{\tmap(\unit)}$ (Russo-Dye). As a consequence, the spectral radius $r(\tmap)$ of a positive unital map $\tmap$ is $1$.

A projection $p\in\A$ \ki{reduces} the positive map $\tmap$ if $\tmap$ leaves the subalgebra $p\A p$ invariant, or equivalently if there exists $t\in\R_{\mathrm{+}}$ such that $\tmap(p)\le tp$. If no non-trivial projection reduces $\tmap$ then it is called 
 \ki{irreducible}. A positive map $\tmap$ is \ki{primitive} if $\tmap^n>0$ for some $n\in\N$. 
As it was shown in \cite[Lemma 2.1]{EHK}, the following holds:
\begin{Lemma}\label{irredpr}
A positive map $\tmap$ is irreducible if and only if $\id+\tmap$ is primitive. 
\end{Lemma}

The following lemma is a key observation in proving LDP for e.g. classical Markov chains (\cite[Section 3.1)]{DZ}):
\begin{Lemma}\label{convergence}
If $\tmap$ is a non-zero positive map with a strictly positive eigenvector then the corresponding eigenvalue is the spectral radius $r:=r(\tmap)$ which is strictly positive, and 
\begin{equation*}
\frac{1}{n}\,\log\,\vfi\bz \tmap^n(x) \jz\xrightarrow[n\to\infty]{} \log\,r
\end{equation*}
for any non-zero positive linear functional $\vfi$ and strictly positive $x\in\A$.
\end{Lemma}
\begin{proof}
Let $\tmap(z)=tz$ for some $0\le t\in\R$ and $0<z\in\A$. For any $x\in\A_{\mathrm{+}}$ we can find $0<\beta_x\in\R$ such that $x\le \beta_x\, z$, hence if $t=0$ then $\tmap\bz\A_{\mathrm{+}}\jz=\{0\}$, i.e. $\tmap$ is the zero map. Thus $0<t$ and $\frac{1}{t}\tmap$ is similar to the unital map $\Psi:\,a\mapsto z^{-1/2}\,\frac{1}{t}\tmap\bz z^{1/2}az^{1/2} \jz\,z^{-1/2}$, hence $1=r(\Psi)=\frac{1}{t}\,r(\tmap)$, i.e.~$t=r$.

If $0<x\in\A$ then we can also find $0<\alpha_x\in\R$ such that $\alpha_x\,z\le x$, hence
\begin{equation*}
\alpha_x\,r^n\,z\le\alpha_x\,\tmap^n(z)\le\tmap^n(x)\le \beta_x\,\tmap^n(z)=\beta_x\,r^n\,z\,,
\end{equation*}
that yields the desired statement.
\end{proof}

As the classical Perron-Frobenius theorem shows, positivity has strong implications on the spectral properties of linear operators. The following fundamental extension of the Perron-Frobenius theorem to $C^*$-algebras was proven in \cite{EHK}:
\begin{thm}\label{PF}
Let $\tmap$ be an irreducible positive map on a finite-dimensional $C^*$-algebra $\A$. 
Then the spectral radius $r$ of $\tmap$ is an eigenvalue of $\tmap$ with geometric multiplicity $1$ and there exists $z>0$ such that $\tmap(z)=rz$. 
\end{thm}
Note that if $\tmap$ is an irreducible positive map then so is $\tmap^*$ (the adjoint taken with respect to the Hilbert-Schmidt inner product $\inner{a}{b}:=\Tr a^*b$), hence there exists $0<\D{\rho}\in\A$ such that $\tmap^*\bz\D{\rho}\jz=r\D{\rho}$, where $r=r(\tmap^*)=r(\tmap)$. 
We can assume that $\Tr\D{\rho}=1$, i.e. $\D{\rho}$ is the density operator of a faithful state $\rho$ with the property $\rho\circ\tmap=r\rho$. As the following lemma shows, irreducibility is also necessary for the above spectral properties:
\begin{Lemma}\label{PF2}
Let $\tmap$ be a positive map on a finite dimensional $C^*$-algebra $\A$. If both $\tmap$ and $\tmap^*$ have strictly positive eigenvectors with geometric multiplicity $1$ then $\tmap$ is irreducible.
\end{Lemma}
\begin{proof}
Let $z$ and $\D{\rho}$ be strictly positive eigenvectors of $\tmap$ and $\tmap^*$, respectively, with a normalization $\inner{\D{\rho}}{z}=1$. As we have seen in Lemma \ref{convergence}, the corresponding eigenvalue is $r:=r(\tmap)$. Obviously $r(\tmapp)=1$ for $\tmapp:=(\id+\tmap)/(1+r)$, and $\tmapp(z)=z,\s\tmapp^*\D{\rho}=\D{\rho}$ hold. Moreover, $1$ is the only eigenvalue of $\tmapp$ on the unit circle. 

We define $P:=\diad{z}{\D{\rho}}$ and follow the argument of \cite[Lemma I.3.3]{Sch}. A straightforward computation shows that $P=P^2$ is a projection onto $\ker\{\id-\tmapp\}$ and $P\tmapp=\tmapp P=P$, hence $P\tilde\tmapp=\tilde\tmapp P=0$ for $\tilde\tmapp:=\tmapp-P$.
If $\alpha$ is a non-zero eigenvalue of $\tilde\tmapp$ with eigenvector $x\in\A$ then $Px=P\bz (1/\alpha)\,\tilde\tmapp(x)\jz=0$, hence $\tmapp(x)=\tilde\tmapp(x)+Px=\alpha x$, i.e. $\alpha$ is an eigenvalue of $\tmapp$ as well. If $\tilde\tmapp(x)=x$ for some $x\in\A$ then the same computation yields that $Px=0$ and $\tmapp x=x$, but the latter implies $Px=x$, therefore $x=0$, i.e. $1$ is not an eigenvalue of $\tilde\tmapp$. As a consequence, $r(\tilde\tmapp)<1$, thus
\begin{equation*}
\frac{1}{(1+r)^n}\,(\id+\tmap)^n=\bz\tmapp\jz^n=(\tilde\tmapp)^n+P\xrightarrow[n\to\infty]{}P\, .
\end{equation*}
As $P$ is positivity increasing, we get $0<(\id+\tmap)^n$ for large enough $n$'s and Lemma \ref{irredpr} gives the irreducibility of $\tmap$.
\end{proof}


\begin{thebibliography}{99}
    \addcontentsline{toc}{chapter}{Bibliography}

\bibitem{Accardi} L.~Accardi: \kii{Non commutative Markov chains}; 
                   \kiii{Proc.~School of Math.~Phys.~Camerino}, 268--295, (1974)

\bibitem{AccF} L.~Accardi, A.~Frigerio: 
          \kii{Markovian cocycles}; 
          \kiii{Proc.~Roy.~Irish Acad.} \textbf{83A}(2), 251--263, (1983)

\bibitem{Araki} H.~Araki: \kii{Gibbs States of a one dimensional quantum lattice};
                           \kiii{Commun.~Math.~Phys.} \textbf{14}, 120--157, (1969)

\bibitem{Ar}
H.~Araki, \kii{On uniqueness of KMS states of one-dimensional quantum lattice systems};
\kiii{Commun.~Math.~Phys.} {\bf 44}, 1--7, (1975)

\bibitem{Araki2} H.~Araki: \kii{Positive cone, Radon-Nikodym theorems, relative Hamiltonian and the Gibbs condition in statistical mechanics. An application of Tomita-Takesaki theory}; in
   \kiii{Proc.~Internat.~School of Physics (Enrico Fermi)}, 64--100, (1976)

\bibitem{ArIon} H.~Araki and P.D.F.~Ion:
            \kii{On the equivalence of KMS and Gibbs conditions for states of quantum lattice systems} 
            \kiii{Commun.~Math.~Phys.} {\bf 35}, 1--12, (1974)

\bibitem{Aud} K.M.R.~Audenaert, J.~Calsamiglia, Ll.~Masanes, R.~Munoz-Tapia, A.~Acin, E.~Bagan, F.~Verstraete.: 
            \kii{Discriminating states: the quantum Chernoff bound};  
           \kiii{Phys.~Rev.~Lett.} \textbf{98} 160501, (2007)
 
\bibitem{BLP} M.~van den Berg, J.T.~Lewis, J.V.~Pule: 
         \kii{The large deviation principle and some models of an interacting boson gas}; 
         \kiii{Commun.~Math.~Phys.} \textbf{118}, 61--85, (1988)

\bibitem{Bhatia} R.~Bhatia: \kii{Matrix Analysis};
              \kiii{Springer}, (1997)

\bibitem{Bjelakovic2} I.~Bjelakovic, J.-D.~Deuschel, T.~Kr\"uger, R.~Seiler, Ra.~Siegmund-Schultze, A.~Szko\l a:
                  \kii{A quantum version of Sanov's theorem};
                  \kiii{Commun.~Math.~Phys.} \textbf{260}, 659--671, (2005)

\bibitem{Bjelakovic} I.~Bjelakovic, J.-D.~Deuschel, T.~Kr\"uger, R.~Seiler, Ra.~Siegmund-Schultze, A.~Szko\l a:
                  \kii{Typical support and Sanov large deviations of correlated states};
                  \kiii{Preprint; math/073772}, (2007)
 
\bibitem{BMV}
D.~Bessis, P.~Moussa and M.~Villani: \kii{Monotonic converging variational approximations
to the functional integrals in quantum statistical mechanics};
\kiii{J.~Math.~Phys.} {\bf 16}, 2318--2325, (1975)

\bibitem{BR2} O.~Bratteli, D.W.~Robinson:
                \kii{Operator Algebras and Quantum Statistical Mechanics 2. (Second ed.)};
                Springer, (1997)

\bibitem{Choi} M.D.~Choi: \kii{Completely positive linear maps on complex matrices};
            \kiii{Linear~Algebra Appl.} \textbf{10}, 285--290, (1975)

\bibitem{DZ} A.~Dembo, O.~Zeitouni: \kii{Large Deviations Techniques and Applications. (Second ed.)};
              \kiii{Springer, Application of Mathematics, Vol.} \textbf{38}, (1998)

\bibitem{EHK} D.E.~Evans, R.~Hoegh-Krohn: \kii{Spectral properties of positive maps on $C^*$-algebras};
               \kiii{J. London Math. Soc.} \textbf{17}, 345--355, (1978)
 
 \bibitem{FNW} M.~Fannes, B.~Nachtergaele, R.F.~Werner:
               \kii{Finitely correlated states on quantum spin chains};
               \kiii{Commun.~Math.~Physics} \textbf{144}, 443--490, (1992)

\bibitem{GLM} G.~Gallavotti, J.L.~Lebowitz, V.~Mastropietro: 
             \kii{Large deviations in rarefied quantum gases};
             \kiii{J.~Stat.~Phys.} \textbf{108}, 831--861, (2002)

 \bibitem{HP} F.~Hiai, D.~Petz: \kii{Entropy densities for algebraic states};
              \kiii{J.~Funct.~Anal.} \textbf{125}, 287--308, (1994)

\bibitem{HMOP} F.~Hiai, M.~Mosonyi, H.~Ohno, D.~Petz: \kii{Free energy density for mean field perturbation of states of a one-dimensional spin chain}; \kiii{Preprint; arXiv:~0706.4148}, (2007)

\bibitem{HMO} F.~Hiai, M.~Mosonyi, T.~Ogawa: 
             \kii{Error exponents for hypothesis testing on a spin chain}; 
             \kiii{preprint}, (2007)

\bibitem{Israel} R.B.~Israel: \kii{Convexity in the Theory of Lattice Gases};
                         \kiii{Princeton University Press}, (1979)

\bibitem{Kato} T.~Kato: \kii{Perturbation Theory for Linear Operators};
              \kiii{Springer}, (1980)

\bibitem{LRB} M.~Lenci, L.~Rey-Bellet: \kii{Large deviations in quantum lattice systems: one-phase region};
              \kiii{J.~Stat.~Phys.} \textbf{119}, 715--746, (2005)

\bibitem{LLS} J.L.~Lebowitz, M.~Lenci, H.~Spohn:
              \kii{Large deviations for ideal quantum systems};
              \kiii{J.~Math.~Phys.} \textbf{41}, 1224--1243, (2000)

\bibitem{LS}
E.H.~Lieb and R.~Seiringer: \kii{Equivalent forms of the Bessis-Moussa-Villani conjecture};
\kii{J.~Stat.~Phys.} {\bf 115}, 185--190, (2004)


\bibitem{NR} K.~Neto$\check{\text{c}}$ny, F.~Redig: \kii{Large deviations for quantum spin systems};
            \kiii{J.~Stat.~Phys.} \textbf{117}, 521--547, (2004)

 
\bibitem{SzN} M.~Nussbaum,  A.~Szko\l a: \kii{A lower bound of Chernoff type for symmetric quantum hypothesis testing}; \kiii{quant-ph/0607216; to appear in Ann.~Statist.}

\bibitem{Petz} D.~Petz: \kii{Quasi-entropies for finite quantum systems};
                       \kiii{Rep.~Math.~Phys.} \textbf{23}, 57--65, (1986)

\bibitem{PRV} D.~Petz, G.P.~Raggio, A.~Verbeure: \kii{Asymptotics of Varadhan-type and the Gibbs variational principle};
              \kiii{Commun.~Math.~Phys.} \textbf{121}, 271--282, (1989)

\bibitem{Petz2}
D.~Petz: \kii{First steps towards a Donsker and Varadhan theory in
operator algebras}; \kiii{in Quantum Probability and Applications
IV}, Lecture Notes in Math., {\bf 1442}, 311--319, (1990)

 

\bibitem{RS} M.~Reed, B.~Simon:
             \kii{Methods of Modern Mathematical Physics I.}
             \kiii{Academic Press}, (1980)

 
 

 
 \bibitem{Rellich} F.~Rellich: \kii{Perturbation Theory of Eigenvalue Problems};
              \kiii{Gordon and Breach}, (1969)

\bibitem{Sakai} S.~Sakai: \kii{Operator Algebras in Dynamical Systems};
                         \kiii{Cambridge University Press}, (1991)
 
 \bibitem{Sch} H.H.~Schaefer: \kii{Banach Lattices and Positive Operators};
              \kiii{Springer}, (1974)
   
 

               
 \end{thebibliography}
\end{document}